\documentclass[epjCONF,columns]{svjour}
\usepackage{graphics}
\usepackage[varg]{txfonts} 
\usepackage[latin1]{inputenc}
\session-title{Hadron Collider Physics Symposium 2011}
\begin{document}
\title{Dibosons from CMS}
\author{Arabella Martelli\inst{1}\inst{2}\fnmsep\thanks{\email{arabella.martelli@cern.ch}} }
\institute{Physics Department,
           Universit\'a degli Studi di Milano Bicocca,\\
           Piazza della Scienza 3, 20156 Milano, Italy    \and
           Laboratoire Leprince-Ringuet,
           Ecole Polytechnique, \\
           91128 Palaiseau Cedex, France}           
\abstract{
It is here presented the diboson production cross section measured by the CMS collaboration
in $pp$ collisions data at $\sqrt{s}$=7 TeV.
$W \gamma$ and $Z \gamma$ results from 2010 analyses (36 pb$^{-1}$) are presented
together with 2011 first measurements of  $WW$, $WZ$ and $ZZ$ final states obtained using 1.1 fb$^{-1}$.
Results obtained with 2010 data are also interpreted in term of anomalous triple gauge couplings.
} 
\maketitle
\section{Introduction}
\label{intro}
The diboson production is a direct probe of the Standard Model.
Its high sensitivity to the self-interaction between gauge bosons 
is a direct consequence of the non-abelian $SU(2) \times U(1)$ gauge symmetry of the SM
and is fully fixed in the SM by the gauge structure of the Lagrangian. 
Any deviation of the SM couplings is an indication of new physics, 
manifested as an increased production cross section for instance.\\
The measurement of triple gauge boson couplings (TGC) is also an important test of the SM as a
useful step to establish major backgrounds to the electroweak searches. 
Moreover, a number of extensions of the SM can manifest themselves in processes with multiple bosons in the final state and 
a measurement of TGCs can be sensitive to new phenomena at high energies, that would require more energy or luminosity to be directly observed.\\
Within the SM, using the effective Lagrangian approach, 
$g_1^V, k^V_1, \lambda_V$ and $h^V_3, h^V_4$ operators are used to describe 
the charged~\cite{Ref:TGCcharged}  and neutral~\cite{Ref:TGCneutral} couplings respectively, 
which are Lorentz and $SU(2) \times U(1)$ invariant and conserve C and P separately.
In the following the 95\% CL intervals for anomalous TGCs are presented, using the HISZ parametrization.
No form-factor is used in the interpretation of the results in order not to make any assumption 
on the energy dependency of new physics that would ensure partial-wave unitarity.\\

A complete description of the detector and informations about objects reconstruction can be found in\cite{Ref:CMS}.
The analyses here presented use data from $pp$ collisions at 7 TeV 
registered by the CMS detector in 2010 (36 pb$^{-1}$) and 2011 (1.1 fb$^{-1}$), focusing on 
the fully leptonic final states reconstructed with high efficiency
over a very wide acceptance. 
Results are presented in the following, with a brief description of the diboson events selection.

\section{$W\gamma$ and $Z\gamma$ production cross section}
\label{sec:1}
$W\gamma$ and $Z\gamma$ diboson processes are studied with analyses sharing similar strategy 
and techniques. $W$ bosons are reconstructed if only one isolated lepton with $p_T>$20 GeV/c 
is found together with a missing transverse energy above 25 GeV/c, while
the $Z$ candidates are selected looking for two isolated leptons with $p_T>$20 GeV/c 
and a dilepton invariant mass above 50 GeV/c$^2$.
For both analyses, only photons with  $p_T>$10 GeV/c are considered. \\
Moreover an angular separation in terms of $\Delta R(\gamma, lepton) = \sqrt{ (\Delta \eta) ^2 + (\Delta \phi)^2}>0.7$ is required.\\
The main backgrounds are due to jets misidentified as photons and are estimated with data driven techniques.
Both $W$+jets and $Z$+jets background contributions are estimated by measuring the 
probability for a jet to be identified as a photon candidate 
in a sample of multi-jet $QCD$ events containing at least one high-quality jet candidate,  
and then folding this probability with the non-isolated photon candidates observed in the $W\gamma$ and $Z \gamma$ samples. \\
Since only electrons and muons final states are considered, 
the fraction of $W \gamma$ events decaying into taus are subtracted as a background, 
once estimated from the simulation (order 3\%).\\

The number of events observed in data and estimated for the backgrounds in each leptonic final state 
is shown in Table~\ref{tab:Wg} for the $W \gamma$ and in Table~\ref{tab:Zg} for the $Z\gamma$ analysis\cite{Ref:WgZg2010}.
No event is observed with more than one photon candidate in the final state.
\begin{table}[!htb]
\caption{Number of events observed in data and background event yield in each leptonic final state
at the end of the $W \gamma$ analysis, for $\int \mathcal{L} \,\, dt$=36 pb$^{-1}$.}
\label{tab:Wg}       
\begin{tabular}{llll}
\hline\noalign{\smallskip}
Final state  & $N_{obs.}$ & $W+jet$                 & other relevant backgrounds \\
\noalign{\smallskip}\hline\noalign{\smallskip}
$e \nu$      &  452       &  220 $\pm$ 16 $\pm$ 14 &  7.7  $\pm$ 0.5  \\
$\mu \nu$    &  520       &  261 $\pm$ 19 $\pm$ 16 &  16.4 $\pm$ 1.   \\
\noalign{\smallskip}\hline
\end{tabular}
\end{table}
\begin{table}[!htb]
\caption{Number of events observed in data and background event yield in each leptonic final state
at the end of the $Z \gamma$ analysis, for $\int \mathcal{L} \,\, dt$=36 pb$^{-1}$.}
\label{tab:Zg}       
\begin{tabular}{lll}
\hline\noalign{\smallskip}
Final state& $N_{obs.}$  & $Z+jet$         \\
\noalign{\smallskip}\hline\noalign{\smallskip}
$e e$      &  81      &  20.5 $\pm$ 1.7 $\pm$ 1.9 \\
$\mu \mu$  &  90      &  27.3 $\pm$ 2.2 $\pm$ 2.3 \\
\noalign{\smallskip}\hline
\end{tabular}
\end{table}

The main systematic uncertainty is due to the estimate of backgrounds
from data and amounts to $\sim$6\% for $W \gamma$ and to $\sim$10\% for $Z \gamma$.
Other significant sources of systematic uncertainties are from the PDF modeling and photon energy scale,
contributing by $\sim$5\% for all final states.\\
The $W \gamma$ cross sections measured in each final state are weighted 
taking into account correlated uncertainties between the two results.
The combined cross sections is 
\begin{eqnarray}
\nonumber \sigma{(pp \to W \gamma + X)}  \times B.R.(W \to l \nu) = \\
\nonumber 56.3 \pm 5.0(\mbox{stat.}) \pm 5.0\mbox{(syst.)} \pm 2.3\mbox{(lumi.)}
\end{eqnarray}

This result agrees well with the NLO prediction~\cite{Ref:Bauer} of 49.4 $\pm$ 3.8 pb.\\
By following the same procedure, the corresponding cross section measured for the $Z \gamma$ production  is
\begin{eqnarray}
\nonumber \sigma{(pp \to Z \gamma + X)} \times B.R.(Z \to ll) = \\
\nonumber 9.4 \pm 1.0 \mbox{(stat.)} \pm 0.6 \mbox{(syst.)} \pm 0.4 \mbox{(lumi.)}
\end{eqnarray}

The theoretical NLO prediction~\cite{Ref:TGCneutral} is 9.6 $\pm$ 0.4 pb, which is in agreement with the measured value.

\section{$WW$ production cross section}
The signature looked for a $WW$ event consists in exactly two well isolated leptons among electrons and muons only,
with $p_T>$20(10) GeV/c for the leading (trailing) lepton, together with a 
significant missing transverse energy ($MET$) to account for the two neutrinos.\\
For this purpose a particular variable is used, namely the missing transverse energy 
is projected along the direction of the closest lepton and its orthogonal component is
used as discriminating variable in case the angle $\Delta \phi(MET, lepton)$ is smaller than $\pi /2$.
This ``projected-$MET$'' is particularly suitable to reject eventual $Z \to \tau \tau$ decays, in case of boosted $Z$.
It is required above 40 GeV for the same flavor $W$ decays ($ee, \mu\mu$) and above 20 GeV for the $e\mu$ channels.\\
Some further cuts are required to reduce the  contamination from background processes with jets, 
heavy hadrons and mis-reconstructed diboson channels.
Events are rejected if containing jets with $p_T>$30 GeV/c, a further veto is applied on ``top-tagged'' jets  
accounting also for soft muons from $b$-quark decays.
To further reduce the Drell-Yan background, events with a dilepton $e^+e^-$, $\mu^+ \mu^-$ 
invariant mass within 15 GeV/c$^2$ of the $Z$ are vetoed. 
Finally the angle in the transverse plane between the dilepton system and the most energetic jet with $p_T>$15 GeV/c 
is required to be smaller than 165 degrees, in the $ee$/$\mu\mu$ final states,
to cope with $Z$ events where the boson recoils against a jet.\\
The main backgrounds are estimated directly from data.
The fake rate measured on jets enriched data samples allows 
to estimate $W$+jets and $QCD$ multi-jet events.
The remaining top background is estimated from data as well by using ``top-tagged'' events 
and applying the corresponding tagging efficiency, which is measured in a data control sample.
The residual $Z$ boson contribution to the $e^+e^-$ and $\mu^+ \mu^-$ final states are estimated
by normalizing the simulation to the observed number of events inside the $Z$ mass window in data.\\

The number of expected signal and background events, for the processes controlled with the simulation 
and with data-driven techniques are reported in Table~\ref{tab:WW}\cite{Ref:ZZWZWW2011}.
\begin{table}[!htb]
\caption{Number of events observed in data and expected event yield for signal and background, 
estimated from data and simulation for $\int \mathcal{L} \,\, dt$=1.1 fb$^{-1}$, after applying the $WW$ selection requirements.}
\label{tab:WW}       
\begin{tabular}{ll}
\hline\noalign{\smallskip}
Sample & Yield \\
\noalign{\smallskip}\hline\noalign{\smallskip}
$qq \to W^+ W^-$       & 349.7 $\pm$ 30.3  \\
$gg \to W^+ W^-$       & 17.2  $\pm$ 1.6   \\
$W+jets$               & 106.9 $\pm$ 38.9  \\
$t\bar{t} + tW$        & 63.8  $\pm$ 15.9  \\
$Z/ \gamma^* \to ll$ + $WZ$ + $ZZ$  & 12.2 $\pm$ 5.3 \\
$Z / \gamma^* \to \tau \tau$            & 1.6 $\pm$ 0.4  \\
$WZ / ZZ$ not in $Z / \gamma^* \to ll$  & 8.5 $\pm$ 0.9  \\
$W + \gamma$           & 8.7 $\pm$ 1.7  \\
$Signal + Background$  & 568.6 $\pm$ 52.2   \\
Data                   & 626  \\
\noalign{\smallskip}\hline
\end{tabular}
\end{table}

\begin{figure}[!htb]
\resizebox{0.75\columnwidth}{!}{%
  \includegraphics{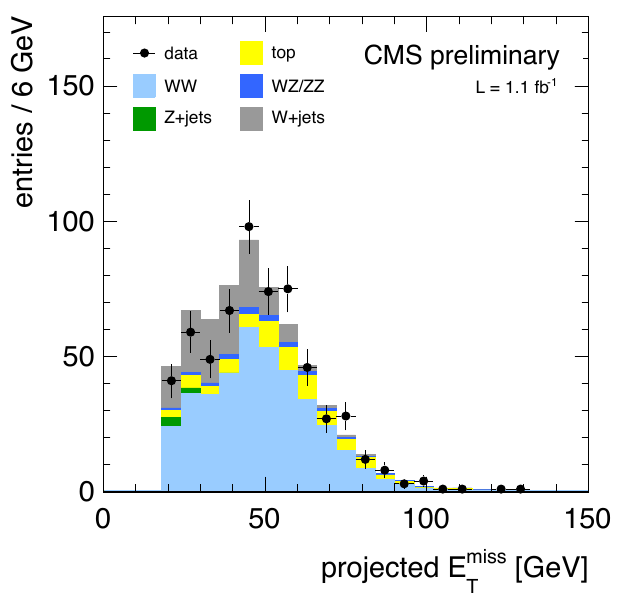} }
\caption{Projected-$MET$ spectrum shown for data and expectation for $\int \mathcal{L} \,\, dt$=1.1 fb$^{-1}$, 
for the events surviving the $WW$ selection.}
\label{fig:WW}     
\end{figure}

The spectrum of the projected-$MET$ at the end of the event selection is shown in Figure~\ref{fig:WW}.

The major systematic uncertainty is from the background estimation from data (order 20\%), 
while $\sim$ 8\% accounts for the signal efficiency.
The signal acceptance corresponding to the selection described is order 70\% of the total phase space.
By using the $W \to l \nu$ branching fraction, the $WW$ production cross section is measured to be
\begin{eqnarray}
\begin{footnotesize}
\nonumber \sigma{(pp \to WW + X)} = 55.3 \pm 3.3 \mbox{(stat.)} \pm 6.9 \mbox{(syst.)} \pm 3.3 \mbox{(lumi.)} 
\end{footnotesize}
\end{eqnarray}
This measurement is consistent with the SM expectation of 43$\pm$2 pb at NLO~\cite{Ref:MCFM} within one standard deviation.

\section{$WZ$ production cross section}
\begin{figure}[!htb]
\resizebox{0.75\columnwidth}{!}{%
  \includegraphics{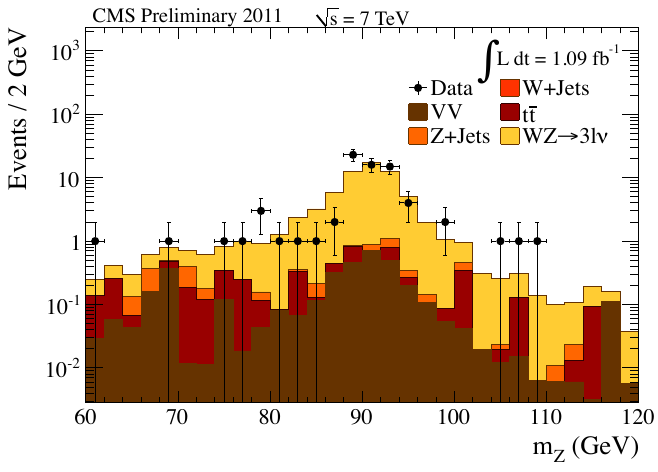} }
\caption{Dilepton invariant mass for the events passing the full $WZ$ selection, 
for $\int \mathcal{L} \,\, dt$=1.1 fb$^{-1}$.}
\label{fig:WZ}     
\end{figure}

In the $WZ \to l' \nu l^+l^-$ channel, three isolated leptons are looked for considering electrons and muons only.
Leptons candidates from $Z$ decays are first selected with $p_T>$20,10 GeV/c for electrons and $p_T>$15,15 GeV/c for muons and 
the $Z$ boson is chosen as the best invariant mass candidate within the range [60, 120] GeV/c$^2$.
Then a third lepton satisfying tight identification and isolation  criteria and with $p_T>$20 GeV/c is required as well as 
a significant missing transverse energy (above 30 GeV) associated to the neutrino 
so to select events containing also a $W$ boson. Events with a second $Z$ candidate reconstructed are rejected. \\
The dilepton invariant mass distribution for 
events surviving the signal selection is presented in Figure~\ref{fig:WZ} for simulation and data.

$Z$+jets and $t\bar{t}$ are estimated from data with the ``matrix-method''.
This technique uses a tag-and-probe procedure to measure the lepton selection efficiency 
and the probability for a jet to fake a lepton, using enriched $Z$ and $Z$+jets data samples.
$ZZ$ and $Z\gamma$ are backgrounds in case of leptons not reconstructed or photon conversions and are 
controlled with the simulation. 
$WZ$ events where bosons decay into tau lepton(s) are considered as background 
and the fraction of such events as estimated from the simulation (order 6\%) 
is subtracted to the same number of events observed in data. \\
The event counts observed in data are reported in Table~\ref{tab:WZ} together 
with the estimated background and signal event yield\cite{Ref:ZZWZWW2011}.
\begin{table}[!htb]
\caption{Number of observed events for the individual final states in the $WZ$ analysis. 
The overall background event yield and the expectation for the signal are also shown.
Numbers correspond to $\int \mathcal{L} \,\, dt$=1.1 fb$^{-1}$.}
\label{tab:WZ}       
\begin{tabular}{llll}
\hline\noalign{\smallskip}
Final state & $N_{obs.}$ & $N^{backg.}_{est.}$ & $N^{WZ}_{exp.}$ \\
\noalign{\smallskip}\hline\noalign{\smallskip}
$eee$       &  22   &  2.98 $\pm$ 0.78   &  14.47 $\pm$ 0.28 \\
$ee \mu$    &  20   &  3.63 $\pm$ 0.87   &  17.4  $\pm$ 0.31 \\
$\mu \mu e$ &  13   &  2.03 $\pm$ 0.58   &  13.95 $\pm$ 0.28 \\
$\mu\mu\mu$ &  10   &  3.15 $\pm$ 0.76   &  18.56 $\pm$ 0.32 \\
\noalign{\smallskip}\hline
\end{tabular}
\end{table}

The signal acceptance corresponding to the selection described is order 50\% of the total phase space.
Concerning the systematic uncertainty, major sources are from the lepton selection
and the background control. In particular a 20\% systematic uncertainty is assigned to $ZZ$ and $Z\gamma$ 
processes, which are estimated from the simulation being minor backgrounds,   
while a systematic uncertainty of $\sim$5\% is assigned to the processes controlled from data.\\
The cross sections measured channel by channel within the $Z$ boson mass range [60, 120] GeV/c$^2$ are 
reported in Table~\ref{tab:WZ2}.
\begin{table}[!htb]
\caption{$WZ$ cross sections for $\int \mathcal{L} \,\, dt$=1.1 fb$^{-1}$ per channel.}
\label{tab:WZ2}       
\begin{tabular}{ll}
\hline\noalign{\smallskip}
Final state & cross section (pb) \\
\noalign{\smallskip}\hline\noalign{\smallskip}
$eee$       &  0.086 $\pm$ 0.022(stat) $\pm$ 0.007(syst) $\pm$ 0.005(lumi) \\
$ee \mu$    &  0.060 $\pm$ 0.017(stat) $\pm$ 0.005(syst) $\pm$ 0.004(lumi) \\
$\mu \mu e$ &  0.053 $\pm$ 0.018(stat) $\pm$ 0.004(syst) $\pm$ 0.003(lumi) \\
$\mu\mu\mu$ &  0.060 $\pm$ 0.016(stat) $\pm$ 0.004(syst) $\pm$ 0.004(lumi) \\
\noalign{\smallskip}\hline
\end{tabular}
\end{table}
The inclusive $WZ$ cross section is computed as a weighted mean taking into account the
correlated systematic uncertainties and  using the $W$ and $Z$ bosons branching ratios. 
It is measured to be
\begin{eqnarray}
\begin{footnotesize}
\nonumber \sigma{(pp \to WZ + X)} = 17.0 \pm 2.4 \mbox{(stat.)} \pm 1.1 \mbox{(syst.)} \pm 1.0 \mbox{(lumi.)} 
\end{footnotesize}
\end{eqnarray}
consistent with the Standard Model prediction~\cite{Ref:MCFM}. 

\section{$ZZ$ production cross section}
The $ZZ$ final state is reconstructed out of two pairs of same flavor, 
opposite charge, isolated leptons, using electrons, muons and taus.
To select the first $Z$ candidate electrons and muons only are considered, if having a $p_T>$20(10) GeV/c for the leading(trailing) lepton 
and invariant mass within [60, 120] GeV/c$^2$, while also taus are taken into account to look for the second $Z$.
A  $p_T>$7(5) GeV/c is required for electrons (muons). 
Different criteria are used to select taus, namely $p_T>$10 GeV/c for leptonic tau decays, $p_T>$20 GeV/c for hadronic taus decays.
Moreover, in presence of taus, it's the  visible mass of the $Z$ boson to be required in the range [30, 80] GeV/c$^2$.
Leptons are required to be isolated, the isolation being measured based on the combination of tracker, ECAL and HCAL measurements.
Finally, a cut is applied on the lepton impact parameter for selected leptons.\\
Leptons from misidentified jets or heavy hadrons decay are a main source of remaining backgrounds.
Concerning final states with taus, also $WZ$ can be a relevant background and it is estimated from  simulations.
$Zbb$ in particular and $t\bar{t}$ are estimated from data, by means of a control region defined by reverting the cut on the 
leptons impact parameter. $Z$+ jets is controlled by measuring the rate of jets faking electrons, muons
and taus respectively. For this purpose a data sample enriched in background is selected, by requiring a $Z$ candidate 
as for the signal plus a pair of same flavor same sign leptons, without isolation criteria.\\

\begin{figure}[!htb]
\resizebox{0.75\columnwidth}{!}{%
  \includegraphics{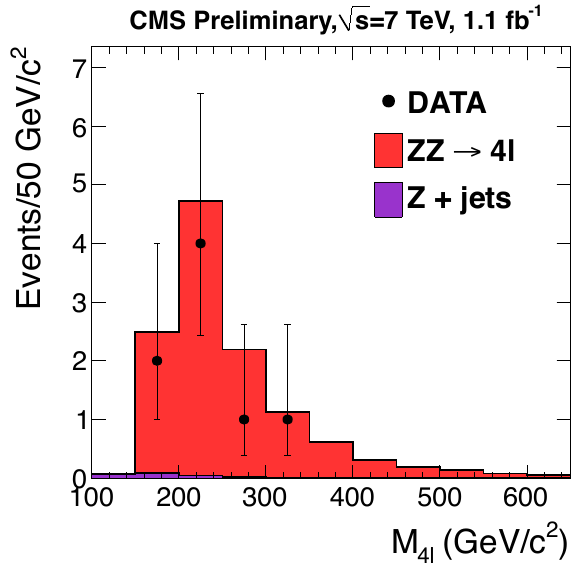} }
\caption{Distribution of the four-lepton reconstructed mass for the final states with 
electrons and muons at the end of the $ZZ$ event selection.
Data and expectation are shown for $\int \mathcal{L} \,\, dt$=1.1 fb$^{-1}$.}
\label{fig:ZZ}     
\end{figure}

The spectrum of the four-lepton invariant mass at the end of the event selection is shown in Figure~\ref{fig:ZZ}
for the final states with electrons and muons only.

In Table~\ref{tab:ZZ} the number of events observed in data, estimated for the backgrounds and as expected for the signal 
are reported\cite{Ref:ZZWZWW2011}.
\begin{table}[!htb]
\caption{Number of expected and observed events for the individual final states in the $ZZ$ analysis, 
as well as the number of background events estimated from data, for $\int \mathcal{L} \,\, dt$=1.1 fb$^{-1}$.}
\label{tab:ZZ}       
\begin{tabular}{llll}
\hline\noalign{\smallskip}
Final state & $N_{obs.}$ & $N^{backg.}_{est.}$ & $N^{ZZ}_{exp.}$ \\
\noalign{\smallskip}\hline\noalign{\smallskip}
$4 \mu$      &  2   &  0.004 $\pm$ 0.004   &  3.7 $\pm$ 0.4 \\
$4 e$        &  0   &  0.14 $\pm$ 0.06     &  2.5 $\pm$ 0.2 \\
$2e 2 \mu$   &  6   &  0.15 $\pm$ 0.06     &  6.3 $\pm$ 0.6 \\
$2 l 2 \tau$ &  1   &  0.8 $\pm$ 0.1       &  1.4 $\pm$ 0.1 \\
\noalign{\smallskip}\hline
\end{tabular}
\end{table}

The presented selection reduces the full signal phase space to order 60\% for the $e,\mu$ and to $\sim$20\% for tau channels respectively.
Main sources of uncertainties are from the background control through data driven techniques and 
the selection of the leptons. As an example, a  3\% systematic error is assigned to the  $e,\mu$ reconstruction, 
while a 6\% to taus.\\
To include all final states in the cross section calculation a simultaneous constrained fit 
on the number of observed events in all decay channels was performed, taking 
into account the systematic uncertainties found, by means of a profile likelihood. 
The resulting $ZZ$ production cross section 
for a pair of $Z$ bosons in the mass range [60, 120] GeV/c$^2$ is measured to be
\begin{eqnarray}
\begin{footnotesize}
\nonumber \sigma{(pp \to Z Z + X)}  = 3.8^{+1.5}_{-1.2} \mbox{(stat.)} \pm 0.2 \mbox{(syst.)} \pm 0.2 \mbox{(lumi.)} 
\end{footnotesize}
\end{eqnarray}
To be compared to the Standard Model NLO prediction of 6.4 $\pm$0.6 pb~\cite{Ref:MCFM}.

\section{Limits on gauge couplings}
Limits on triple gauge couplings were set using 36 pb$^{-1}$ of data from 2010 $pp$ collisions at 7 TeV.\\
To measure such parameters, profile likelihood fits are performed on a relevant spectrum, 
taking the SM prediction as reference for comparison with the measurement. 
This procedure allows to quantify eventual anomalous coupling parameters, which would bring to an  
enhancement of the diboson production cross section, in particular at high boson transverse momentum, 
if introduced in the SM Lagrangian.
The presented 95\% C.L. intervals for the measured TGCs are obtained by varying one of the couplings, 
while fixing the remaining ones to the SM values. \\

Within the $W \gamma$ and $Z \gamma$ analyses described in section~\ref{sec:1},  
$WW \gamma$, $ZZ \gamma$ and  $Z \gamma \gamma$ were measured, by looking at the spectrum of the photon transverse energy\cite{Ref:WgZg2010}.
The results presented in Table~\ref{tab:TGC1} already allow for a good sensitivity to the neutral anomalous couplings.
\begin{table}[!htb]
\caption{95\% C.L. limits on  $WW \gamma$, $ZZ \gamma$ and  $Z \gamma \gamma$ couplings at $\sqrt{s}$ = 7 TeV,
for $\int \mathcal{L} \,\, dt$=36 pb$^{-1}$. These are complementary to the previous results on vector boson self-interactions at lower energy.}
\label{tab:TGC1}       
\begin{tabular}{lll}
\hline\noalign{\smallskip}
Coupling                                       & \multicolumn{2}{c}{ Parameters}         \\
\noalign{\smallskip}\hline\noalign{\smallskip}
    $WW \gamma$                       &       -1.11  $ < \Delta k_{\gamma} < $ 1.04    &  -0.18 $ <{\lambda}_{\gamma}<$ 0.17    \\
     $ZZ \gamma$                        & -0.07, $ < h_3^{Z} <$  0.07                                 &         -0.0005, $ < h_4^{Z} <$  0.0006    \\
      $Z \gamma \gamma$          &      -0.05, $ < h_3^{\gamma} <$  0.06                  &     -0.0005, $ < h_4^{\gamma} <$  0.0005      \\
\noalign{\smallskip}\hline
\end{tabular}
\end{table}

Also the $WW$ analysis here presented was already performed in 2010 providing limits on the  
$WW \gamma$, $WW Z$ couplings. For this purpose the discriminating spectrum used was the leading lepton $p_T$\cite{Ref:WW2010}.
Results are shown in Table~\ref{tab:TGC2}.
\begin{table}[!htb]
\caption{95\% C.L. limits on one-dimensional fit results for anomalous TGC obtained within the $WW$ analysis  
for $\int \mathcal{L} \,\, dt$=36 pb$^{-1}$ of 2010 data at 7 TeV.}
\label{tab:TGC2}       
\begin{tabular}{lll}
\hline\noalign{\smallskip}
Coupling                                       & \multicolumn{2}{c}{ Parameters}         \\
\noalign{\smallskip}\hline\noalign{\smallskip}
$WW \gamma$                          &   -0.61 $< \Delta k_{\gamma} < $ 0.65     &                                                     \\
$WWZ$                          &  -0.19 $< \lambda_Z < $ 0.19                   &  -0.29  $< \Delta g_{Z} < $ 0.31                  \\
\noalign{\smallskip}\hline
\end{tabular}
\end{table}


\begin{thebibliography}{}
\bibitem{Ref:WgZg2010}
The CMS collaboration, \textit{Measurement of $W\gamma$ and $Z\gamma$ production in $pp$ collisions at $\sqrt{s}$=7 TeV}, 
Phys. Lett. \textbf{B701}, (2011) 535-555, doi:10.1016/j.physletb.2011.06.034
\bibitem{Ref:ZZWZWW2011}
The CMS collaboration, \textit{Measurement of $WW$ and observation of $WZ$ and $ZZ$ in leptonic modes}, 
CMS-PAS \textbf{EWK-11-010}, (2011)
\bibitem{Ref:WW2010}
The CMS collaboration, \textit{Measurement of $W^+W^-$ production and search for Higgs boson in $pp$ collision at $\sqrt{s}$=7 TeV}, 
Phys. Lett. \textbf{B699}, (2011) 25-47, doi:10.1016/j.physletb.2011.03.96
\bibitem{Ref:CMS}
The CMS collaboration, \textit{The CMS experiment at the CERN LHC}, 
JINST \textbf{3}, (2008) S08004, doi:10.1088/1748-0221/3/08/S08004
\bibitem{Ref:TGCcharged}
K. Hagiwara and R.D. Peccei and D. Zeppenfeld, \textit{Probing the Weak Boson Sector in $e^+e^- \to W^+ W^-$}, 
Nucl. Phys. \textbf{B282}, (1987) 253
\bibitem{Ref:TGCneutral}
U. Baur and E. Berger, \textit{Probing the weak-boson sector in $Z \gamma$ production at hadron colliders}, 
Phys. Rev. \textbf{D 47}, (1993) 4889
\bibitem{Ref:MCFM}
C. W. John Campbell and Keith Ellis, \textit{MCFM - Monte Carlo for FeMtobarn processes}
\bibitem{Ref:Bauer}
U. Baur and T. Han,and J. Ohnemus, \textit{QCD corrections to hadronic $W\gamma$ production with nonstandard $WW \gamma$ couplings}, 
Phys. Rev. \textbf{D48}, (1993) 5140
\end{thebibliography}
\end{document}